\documentclass{article}

\usepackage[english]{babel}

\usepackage[letterpaper,top=2cm,bottom=2cm,left=3cm,right=3cm,marginparwidth=1.75cm]{geometry}

\usepackage{amsmath}
\usepackage{amssymb}
\usepackage{graphicx}
\usepackage[colorlinks=true, allcolors=blue]{hyperref}
\usepackage{enumerate} 

\usepackage{booktabs} 
\usepackage{multirow} 
\usepackage{siunitx} 

\usepackage[title]{appendix}
\usepackage{multicol}
\usepackage[section]{placeins}

\usepackage{authblk}

\title{Effective dark matter component presents a robust signature of negative pressure by the DESI observations}
\author[ 1 ]{Hao Xu \thanks{Email: hao\_xu@mail.nankai.edu.cn}}
\author[ 1,2 ]{Xinhe Meng \thanks{Email: xhm@nankai.edu.cn, corr.}}
\affil[1]{School of Physics, Nankai University, Tianjin 300071, P.R.China}
\affil[2]{ITP-CAS, Beijing 100190, P.R.China}

\date{}

\begin{document}
\maketitle
\begin{abstract}
 Comprehensive cosmological analysis of an effective non-standard dark matter (NSDM) model, characterized by an equation of state $w_{\mathrm{dm}} = w_2 a^2$, which allows for mild deviations from the previously assumed pressureless cold dark matter, is elaborated in the present work. This effective description framework is the scenarios that matter contents coupled to three distinct single-parameter dynamical dark energy models: i.e, the thawing scalar field, the Modified Emergent Dark Energy (MEDE) scenario, and the constant-$w$ model. We constrain these frameworks by using the latest cosmological probes, including the Planck 2018 Cosmic Microwave Background (CMB) distance priors, the Baryon Acoustic Oscillation (BAO) measurements from the Data Release 2 of the Dark Energy Spectroscopic Instrument (DESI), and three compilations of Type Ia Supernovae (SN~Ia) namely the Dark Energy Survey Year 5 (DESY5) compilation, the Union3 compilation, and the PantheonPlus (PP) sample. Across all three dark energy scenarios and all dataset combinations, we find a consistent preference for negative values of the parameter $w_2$. Furthermore, this result is robust against the choice of dark energy parametrization, suggesting a model-independent deviation from "standard" cold dark matter. This result indicates that the dark matter fluid possesses a small but non-vanishing negative pressure, meaning a non-cold nature. While the inferred Hubble constant $H_0$ remains consistent with the Planck $\Lambda$CDM value and does not fully alleviate the $H_0$ tension with local measurements, the persistent detection of $w_2 < 0$ across a wide range of independent cosmological probes provides compelling evidence for new physics in the dark matter sector---suggesting that dark matter may be better described as an effective fluid endowed with a mild negative pressure, rather than as a perfectly cold, pressureless substance.
\end{abstract}

\section{Introduction}\label{sec:Introduction}	
    The $\Lambda$CDM model, the so called standard cosmological framework based on General Relativity, enjoys robust supports from multiple independent cosmological probes \cite{Planck2018-VI,riess1998,perlmutter1999,alam2021,ACT2022,review_LCDM_weinberg}, while the framework of the $\Lambda$CDM model includes two enigmatic constituents: mainly cold dark matter (DM) and dark energy (DE). Observational evidence for dark matter arises from galactic rotation curves \cite{Rubin1970,rubin1980}, gravitational lensing \cite{tyson1990,clowe2004,clowe2006,massey2007}, anisotropies in the cosmic microwave background \cite{Planck2018-VI}, and the statistical distribution of large-scale structure \cite{alam2021,eisenstein2005,DESI2024bao} mostly. Dark energy is primarily inferred from observations of Type Ia supernovae, which has revealed the late-time accelerated expansion of the Universe \cite{riess1998,perlmutter1999,brout2022} within the $\Lambda$CDM model, and dark matter is normally described as a pressureless perfect fluid with the equation-of-state parameter $w_{\rm dm}=0$, while the dark energy is about equivalent to a cosmological constant, simply modeled as a perfect fluid with $w_{\rm de} = -1$.
 
    Despite its considerable successes, the standard cosmological model faces several observational tensions. Issues such as the missing satellite \cite{klypin1999missing,moore1999dark}, the ``too big to fail'' \cite{boylan2012too}, and the core-cusp \cite{moore1999cold,springel2008aquarius} problems hint that dark matter might not be perfectly cold. A rich landscape of theoretical alternatives has been explored to address these tensions and deepen our understanding of the dark sector beyond the standard assumptions of pressureless cold dark matter, including warm DM \cite{Blumenthal1982,Bode2001}, fuzzy DM \cite{Hu2000,Marsh2014}, interacting DM \cite{Spergel2000}, and decaying DM \cite{Wang2014}. 
    Furthermore, the persistent Hubble tension ($H_0$ problem) \cite{2021Snowmass2021,hu2023hubble} and recent BAO measurements from the Dark Energy Spectroscopic Instrument (DESI) \cite{2025DESI,2025Dynamical} suggest a possible departure from a simple cosmological constant, favoring dynamical dark energy models. Similarly as dark matter, numerous dynamical dark energy models have been proposed beyond the cosmological constant, including: Interacting dark energy (IDE) models \cite{2004Can,yang2018interacting,2020Interacting}, where dark matter (DM) and dark energy (DE) share interactions other than gravitational; Early dark energy (EDE) \cite{2013How,2019Early,Tanvi2016Dark} which behaves like $\Lambda$ at $z \ge 3000$ and decays away as radiation or faster at later times; Phenomenologically Emergent Dark Energy (PEDE) \cite{2019A,2020Evidence}; A DE component with an equation of state allowing for deviation from the cosmological constant $\Lambda$, both constant or dynamical with redshift \cite{2019Observational,2019Yang_O}. Although these studies have so far yielded no definitive evidence for non-cold DM and $\Lambda$ DE, they have provided insightful approaches for modifying the standard cosmological model.

    In an early phenomenological study \cite{2004Can}, $\Lambda(t)$CDM model has been adopted to investigate “decaying vacuum cosmology”. Rather than starting from a specific vacuum decay law, such as $\Lambda \propto H^2$ or $\Lambda \propto R$, they assumed that cold dark matter (CDM) no longer dilutes strictly as $\rho_m \propto a^{-3}$, but instead follows $\rho_m \propto a^{-3+\varepsilon}$, where $\varepsilon$ is a small parameter characterizing the deviation induced by vacuum energy decaying into matter. This effective modification implies a non-standard expansion history for dark matter and, implicitly, an effective equation of state $w_{\rm dm} \ne 0$, thereby suggesting the possibility of introducing non-cold dark matter within extended dark energy frameworks. In a previous paper \cite{2025Examining}, a non-standard dark matter (NSDM) model was proposed, in which the equation of state of dark matter is parameterized as $w_{\rm dm} = w_2 a^2$. The $ww_2\text{DM}$ model constructed by replacing cold dark matter in the $w$CDM framework with this NSDM was shown to not only provide a statistically significant signal beyond cold dark matter but also avoid violating the null energy condition, all while being better supported by observations than $\Lambda$CDM. A natural and compelling question arises: Are the appealing properties of this NSDM model robust when it is coupled with different dynamical dark energy sectors? Different parameterizations of dark energy and dark matter can significantly alter the background expansion history and the growth of structure, so DE component potentially affect the constraints on and the behavior of the NSDM component. To address this question systematically, we couple the $w_2$-parameterized NSDM with three distinct, well-motivated, single-parameter dark energy models: a thawing scalar field model, which leads to a specific evolving $w_{\rm de}(a)$ parameterized by its present value $w_0$; the Modified Emergent Dark Energy (MEDE) model, a generalization of the Phenomenologically Emergent Dark Energy scenario, parameterized by $\alpha$; a simple constant-$w$ dark energy model, where $w_{de}$ is a constant different from $-1$. The resulting composite models are named the $w_0 w_2$DM, $\alpha w_2$DM, and $ww_2$DM models, respectively.
    
    This work therefore investigates whether these compelling advantages are maintained when this dark matter model is coupled with other single-parameter dark energy parametrizations. So the aims of this work are to perform a comprehensive comparative analysis. We seek to determine: (i) whether the preference for a non-zero $w_2$ (indicative of non-cold dark matter) persists across different dark energy backgrounds, (ii) how the different DE models influence the constraints on cosmological parameters like $H_0$, and (iii) which combination of NSDM and DE is most favored by the current ensemble of data.We constrain these models using the latest cosmological datasets, including the \textit{Planck} CMB distance priors, the groundbreaking Baryon Acoustic Oscillation (BAO) measurements from the DESI Data Release 2, and three independent Type Ia Supernova (SN~Ia) compilations (PP, Union3, and DES-Y5). We employ Markov Chain Monte Carlo (MCMC) methods to obtain robust posterior distributions for the parameters and perform model comparisons using the difference in minimum $\chi^2$ values.

    The article is structured as follows. In Section~\ref{sec:Model}, we detail the theoretical framework of the three composite models. Section~\ref{sec:Observational Data} describes the observational data and the statistical methodology. The main results, including parameter constraints and model comparisons, are presented and discussed in Section~\ref{sec:results}. Finally, we summarize our conclusions in Section~\ref{sec:conclusion}.

\section{The Cosmological Model}\label{sec:Model}

    We consider a cosmological framework consisting of radiation, baryons, dark matter, and dark energy. We assume a spatially flat universe, which is quantified by the curvature density parameter
    \begin{equation}
        \Omega_k = 0,
        \label{eq:Omega_k}
    \end{equation}
    consistent with current precision cosmological observations. DM and DE are modeled using phenomenological equations of state (EOS):
    \begin{equation}
        p_{\mathrm{dm}} = w_{\mathrm{dm}}(a)\,\rho_{\mathrm{dm}},
        \label{eq:p_dm}
    \end{equation}
    \begin{equation}
        p_{\mathrm{de}} = w_{\mathrm{de}}(a)\,\rho_{\mathrm{de}},
        \label{eq:p_de}
    \end{equation}
    where $ a $ is the scale factor, and $ w_{\mathrm{dm}}(a) $, $ w_{\mathrm{de}}(a) $ are their respective equation-of-state (EoS) parameters.The corresponding density evolutions are obtained from energy conservation, $\dot{\rho} + 3H(\rho + p) = 0$, yielding
     \begin{equation}
        \rho_{\mathrm{dm}}(a) = \rho_{\mathrm{dm},0}\,f_{\mathrm{dm}}(a), \quad \text{with} \quad 
        f_{\mathrm{dm}}(a) \equiv \exp\!\left[ -3 \int_1^a \frac{1 + w_{\mathrm{dm}}(a')}{a'}\,da' \right],
        \label{eq:f_dm}
    \end{equation}
    \begin{equation}
        \rho_{\mathrm{de}}(a) = \rho_{\mathrm{de},0}\,f_{\mathrm{de}}(a), \quad \text{with} \quad 
        f_{\mathrm{de}}(a) \equiv \exp\!\left[ -3 \int_1^a \frac{1 + w_{\mathrm{de}}(a')}{a'}\,da' \right].
        \label{eq:f_de}
    \end{equation}
    Under the flatness condition~\eqref{eq:Omega_k}, the expansion history is governed by the Friedmann equation:
    \begin{equation}
        H^2(a) = H_0^2 \left[ 
            \Omega_{r0}\,a^{-4} + 
            \Omega_{b0}\,a^{-3} + 
            \Omega_{\mathrm{dm},0}\,f_{\mathrm{dm}}(a) + 
            \Omega_{\mathrm{de},0}\,f_{\mathrm{de}}(a)
        \right],
        \label{eq:Hubble}
    \end{equation}
    where $H(a) \equiv \dot{a}/a$ is the Hubble parameter, $H_0$ is its present-day value, and $\Omega_{r0}$, $\Omega_{b0}$, $\Omega_{\mathrm{dm},0}$, and $\Omega_{\mathrm{de},0}$ denote the present-day density parameters for radiation, baryons, dark matter, and dark energy, respectively. The flatness condition implies the normalization constraint
    \begin{equation}
        \Omega_{r0} + \Omega_{b0} + \Omega_{\mathrm{dm},0} + \Omega_{\mathrm{de},0} = 1.
    \end{equation}

    \subsection{Dark Matter Model}
    The study in \cite{2025Examining} proposed a parametrization for the EoS of DM. According to the constraints from CMB observations, the dark matter model approximates the cold dark matter (CDM) scenario as the scale factor $a \to 0$ , such that its equation-of-state parameter satisfies 
    \begin{equation}
        w_{dm}(a)|_{a=0} = 0
        \label{w_dm}
    \end{equation}
    Therefore, the equation of state parameter for dark matter can be Taylor expanded around $a=0$: 
    \begin{equation}
        w_{dm}(a) = \sum_{n=0}^{\infty}w_n a^n,
        \label{eq:w_dm_taylor}
    \end{equation}
    which, together with equation~\eqref{w_dm}, implies
    \begin{equation}
        w_0 = 0 ,
    \label{eq:w0_w1_zero}
    \end{equation}
    For the same reason that dark matter behaved very similarly to CDM in the early universe—due to constraints from CMB observations—the authors assumed that the derivative of the dark matter EoS with respect to the scale factor vanishes at $a = 0$:
    \begin{equation}
        w_1 = \left.\frac{\mathrm{d}w_{\mathrm{dm}}(a)}{\mathrm{d}a}\right|_{a=0} = 0.
    \end{equation}
    To ensure sufficient precision in parameter fitting while minimizing the number of free parameters, only the first non-vanishing term in the Taylor series is retained. This leads to the following parametrization of the dark matter model:
    \begin{equation}
        w_{\mathrm{dm}}(a) = w_2 a^2,
        \label{eq:w_dm_param}
    \end{equation}
    where $w_2$ is a single free parameter characterizing deviations from CDM. The underlying motivation for this parametrization is that, although dark matter behaves very much like CDM, there may exist higher-order corrections that manifest as small deviations from the standard CDM scenario previously unaccounted for. This form, as an effective DM from, guarantees $w_{\mathrm{dm}} \to 0$ as $a \to 0$, satisfying early-universe constraints, while allowing for non-zero pressure at low redshifts.
    
    \subsection{Dark Energy}
    We now specify three physically motivated dark energy models, each introducing one additional free parameter beyond $w_2$. When combined with the DM model above, they define three distinct cosmological scenarios.
    
    \subsubsection{Model 1: Thawing Scalar Field Dark Energy ($w_0 w_2$DM)}
    The study in the article \cite{2013CONSTRAINTS} focuses on a particular class of dark energy models—thawing scalar field models. When the slow-roll conditions,
    \begin{equation}
    \left(\frac{1}{V}\frac{dV}{d\phi}\right)^{2} \ll 1
    \quad \text{and} \quad
    \frac{1}{V}\frac{d^{2}V}{d\phi^{2}} \ll 1,
    \label{eq:slow_roll}
    \end{equation}
    are satisfied, a universal relationship exists between the equation of state parameter $w$ and the dark energy density parameter $\Omega_\phi$, which is independent of the specific form of the potential:
    \begin{equation}
        1+w = (1+w_0)\left[\frac{1}{\sqrt{\Omega_{\phi0}}}-(\Omega_{\phi0}^{-1}-1)\tanh^{-1}\sqrt{\Omega_{\phi0}}\right]^{-2}\left[\frac{1}{\sqrt{\Omega_{\phi}}}-\left(\frac{1}{\Omega_{\phi}}-1\right)\tanh^{-1}(\sqrt{\Omega_{\phi}})\right]^{2}.
        \label{eq:w_universal}
    \end{equation}
    The core concept of the thawing scalar field model is that the equation of state parameter $w$ evolves, starting from $w \approx -1$ in the early universe and gradually `thawing' to deviate from $-1$. Therefore, if $w \approx -1$, the evolution of the dark energy density fraction $\Omega_\phi$ from the standard $\Lambda$CDM model,
    \begin{equation}
        \Omega_{\phi}=\frac{1}{1+(\Omega_{\phi0}^{-1}-1)a^{-3}},
        \label{eq:Omega_phi_LCDM}
    \end{equation}
    is used. This leads to the equation of state parameter satisfying
    \begin{equation}
    \begin{split}
        w(a) = {}& -1 + (1+w_{0})\left[\frac{1}{\sqrt{\Omega_{\phi0}}}-(\Omega_{\phi0}^{-1}-1)\tanh^{-1}\!\sqrt{\Omega_{\phi0}}\right]^{-2} \\
        & \times \left[\sqrt{1+(\Omega_{\phi0}^{-1}-1)a^{-3}} - (\Omega_{\phi0}^{-1}-1)a^{-3}\tanh^{-1}\!\left(\left[1+(\Omega_{\phi0}^{-1}-1)a^{-3}\right]^{-1/2}\right)\right]^2.
    \end{split}
    \label{eq:w_a_thawing}
    \end{equation}
    
    Furthermore, by performing a CPL parametrization of the equation of state parameter around $a=1$,
    \begin{equation}
        w_{de} = w_{0} + w_{a}(1-a),
        \label{eq:CPL}
    \end{equation}
    with
    \begin{equation}
        w_{a} = 6(1+w_{0}) \frac{(\Omega_{de0}^{-1}-1)(\sqrt{\Omega_{de0}}-\tanh^{-1}(\sqrt{\Omega_{de0}}))} {{\Omega_{de0}}^{-\frac{1}{2}}-(\Omega_{de0}^{-1}-1)\tanh^{-1}(\sqrt{\Omega_{de0}})}.
        \label{eq:w_a_expr}
    \end{equation}
    The resulting parameterization for the equation of state contains only a single parameter: $w_0$. This dark energy model, when combined with the aforementioned dark matter model, is referred to as the $w_0 w_2$DM model.
    
    \subsubsection{Model 2: Modified Emergent Dark Energy ($\alpha w_2$DM)}
    The Modified Emergent Dark Energy (MEDE) framework \cite{2021Modified} posits that dark energy emerges dynamically around a critical redshift. The normalized DE density is modeled as:
    \begin{equation}
        \tilde{\Omega}_{\mathrm{DE}}(z) = \Omega_{\mathrm{DE},0} \cdot G(z), \quad \text{with} \quad G(z) = 1 - \tanh\!\left[ \alpha \log_{10}(1+z) \right],
        \label{eq:Omega_DE_MEDE}
    \end{equation}
    where the scale factor $a$ and redshift $z$ are related by 
    \begin{equation}
        a = \frac{1}{1+z} .
    \end{equation}
    Here, \(\alpha\) controls the sharpness and timing of the emergence: \(\alpha = 0\) recovers \(\Lambda\)CDM, while \(\alpha = 1\) corresponds to the original PEDE model. Using energy-momentum conservation, the equation-of-state parameter of dark energy $w_{\mathrm{de}}$ can be expressed as a function of redshift $z$ ,
    \begin{equation}
        w_{\mathrm{de}}(z) = -1 - \frac{\alpha}{3\ln(10)} \left(1 + \tanh\!\left[ \alpha \log_{10}(1+z) \right] \right).
        \label{eq:w_de_MEDE}
    \end{equation}
    Thus, the DE sector is governed by a single parameter \(\alpha\). Coupled with the \(w_2 a^2\) dark matter model, this yields the $\alpha w_2$DM model.

    \subsubsection{Model 3: Constant-$w$ Dark Energy ($w w_2$DM)}
    As a minimal extension of $\Lambda$CDM, the dark energy equation of state parameter $w$ corresponding to the cosmological constant $\Lambda$ in the standard $\Lambda$CDM model is replaced by a constant but not necessarily -1:
    \begin{equation}
        w_{de} = w = \text{constant}.
        \label{eq:w_const}
    \end{equation}
    This introduces a single parameter $w$, reducing to the $\Lambda$CDM model when $w = -1$. Combined with the $w_2 a^2$ dark matter model, this defines the $w w_2$DM model.

    \subsubsection{Summary}
    In summary, our analysis compares three cosmological models, all of which share the same dynamical dark matter sector characterized by $w_{\mathrm{dm}} = w_2 a^2$, but differ in their description of dark energy. Model 1 adopts a thawing scalar field for dark energy and is parameterized by $(w_0, w_2)$. Model 2 implements the early dark energy scenario known as MEDE , with parameters $(\alpha, w_2)$. Model 3 assumes a constant equation-of-state parameter for dark energy, described by $(w, w_2)$. Each model introduces exactly two free parameters beyond the standard $\Lambda$CDM baseline, thereby allowing a fair and consistent comparison of their respective abilities to fit current observational data.
    
    \section{Observational Data And Methodology}\label{sec:Observational Data}
    This section presents the observational datasets employed to constrain the parameters of the three proposed cosmological models, along with the statistical framework used for parameter inference and model comparison.

    \subsection{Data Components}
    \subsubsection{Cosmic Microwave Background}
    All three models are constructed to recover the standard $\Lambda$CDM expansion history at high redshift, ensuring that deviations in the early universe, particularly around recombination, are negligible. Under this assumption, it is justified to use the \textit{Planck} 2018 CMB distance priors from Table F1 of \cite{2025Union} instead of using the full set of temperature anisotropy and polarization power spectrum data. This approach retains the essential geometric information while significantly reducing computational cost.
    
    The distance priors consist of three quantities: the shift parameter $R$, the acoustic angular scale $\theta_*$, and the physical baryon density $\omega_b = \Omega_{b0} h^2$, where $h = H_0 / (100~\mathrm{km\,s^{-1}\,Mpc^{-1}})$. The shift parameter is defined as
    \begin{equation}
        R = \sqrt{\Omega_{m0} H_0^2}\, D_M(z_*),
        \label{eq:shift_R}
    \end{equation}
    with $\Omega_{m0} = \Omega_{b0} + \Omega_{\rm dm,0}$ and $D_M(z_*) = (1+z_*) D_A(z_*) = \int_0^{z_*} \! dz / H(z)$ denoting the comoving angular diameter distance to the recombination redshift $z_*$. The acoustic scale is given by
    \begin{equation}
        \theta_* = \frac{r_s(z_*)}{D_M(z_*)},
        \label{eq:theta_star}
    \end{equation}
    where $r_s(z_*)$ is the sound horizon at recombination,
    \begin{equation}
        r_s(z_*) = \int_{z_*}^{\infty} \frac{c_s(z)}{H(z)}\, dz,
        \label{eq:r_s}
    \end{equation}
    and the sound speed reads $c_s(z) = 1/\sqrt{3(1 + \bar{R}_b/(1+z))}$ with $\bar{R}_b = 3\omega_b / (4 \times 2.469 \times 10^{-5})$. The recombination redshift $z_*$ is approximated using the fitting formula from \cite{1996Small}:
    \begin{equation}
        z_* = 1048\left(1 + 0.00124\,\omega_b^{-0.738}\right)\left(1 + g_1 \omega_m^{g_2}\right),
        \label{eq:z_star}
    \end{equation}
    where
    \begin{equation}
        g_1 = \frac{0.0783\,\omega_b^{-0.238}}{1 + 39.5\,\omega_b^{0.763}}, \quad
        g_2 = \frac{0.56}{1 + 21.1\,\omega_b^{1.81}},
        \label{eq:g1_g2}
    \end{equation}
    and $\omega_m = \Omega_{m0} h^2$, where $h = H_0 / (\qty{100}{km.s^{-1}.Mpc^{-1}})$.
    
    \subsubsection{Baryon Acoustic Oscillations}
    We include the latest BAO measurements from the DESI Data Release 2, which delivers the most precise BAO constraints to date across a broad redshift interval $0.1 < z < 4.2$. DESI DR2 reports the ratios $D_V(z)/r_d$, $D_M(z)/r_d$, and $D_H(z)/r_d$ at seven effective redshifts, where the volume-averaged distance is defined as $D_V(z) = [(1+z)^2 D_M^2(z) \, z / H(z)]^{1/3}$, the Hubble distance as $D_H(z) = 1/H(z)$, and $r_d = r_s(z_d)$ is the sound horizon at the baryon drag epoch $z_d$. The drag redshift is computed via the fitting formula \cite{1996Small}:
    \begin{equation}
        z_d = 1291\, \frac{\omega_m^{0.251}}{1 + 0.659\,\omega_m^{0.828}} \cdot \frac{1 + b_1 \omega_b^{b_2}}{0.962},
        \label{eq:z_d_params}
    \end{equation}
    where
    \begin{equation}
        b_1 = 0.313\,\omega_m^{-0.419} \left(1 + 0.607\,\omega_m^{0.674}\right), \quad
        b_2 = 0.238\,\omega_m^{0.223}.
        \label{eq:b1_b2}
    \end{equation}
    These BAO data provide powerful leverage on the late-time expansion rate and are highly complementary to the CMB.

    \subsubsection{Type Ia Supernovae}
    To further constrain the luminosity distance–redshift relation, we incorporate three independent Type Ia supernova (SN Ia) compilations. The first is the Dark Energy Survey Year 5 (DESY5) sample, which includes 1,635 SN Ia in the redshift range $0.10 < z < 1.13$, augmented by an external low-redshift anchor of 194 SN Ia spanning $0.025 < z < 0.10$ \cite{DES2024SNe}. The second is the Union3 compilation, comprising 2,087 SN Ia over $0.05 < z < 2.26$ \cite{2025Union}. The third is the Pantheon+ (PP) sample, containing 1,550 SN Ia across $0.01 < z < 2.26$ \cite{2021The}. Each dataset is analyzed using its published covariance matrix and expressed in terms of the standardized distance modulus 
    \begin{equation}
        \mu(z) = 5 \log_{10}[d_L(z)/(10~\mathrm{pc})],
    \end{equation}
    where the luminosity distance is $d_L(z) = (1+z) D_M(z)$.

    \subsection{Data Combinations and Statistical Methodology}
    To evaluate the constraining power of different observational probes and test the robustness of our conclusions, we consider four distinct combinations of datasets. The first combination, labeled CMB+DESI, uses only the CMB distance priors and DESI BAO measurements, forming a purely geometric baseline that combines high-redshift information from the CMB at $z \sim 1100$ with late-time BAO measurements spanning $0.1 \lesssim z \lesssim 4.2$. This combination relies exclusively on standard rulers-namely the sound horizon at radiation drag-and is insensitive to the luminosity calibration or astrophysical systematics inherent in supernova observations. The remaining three combinations each augment this geometric backbone with a different SN Ia compilation: CMB+DESI+PP incorporates the Pantheon+ sample, CMB+DESI+Union3 substitutes it with the Union3 compilation, and CMB+DESI+DESY5 employs the DESY5 dataset, which benefits from dedicated photometric calibration and a well-characterized low-redshift anchor. Because SN Ia act as standard candles, their inclusion introduces complementary information on the expansion history at intermediate redshifts ($0.01 \lesssim z \lesssim 2.3$) and breaks geometric degeneracies—particularly those involving the dark energy EOS parameters and the dark matter EOS paremeter $w_2$. Comparing results across these four datasets thus allows us to isolate the impact of the supernova component, assess the consistency among different SN Ia samples, and determine whether any preference for extended models over $\Lambda$CDM is driven primarily by geometric probes, luminosity-distance measurements, or their synergy.
    
    For each model and dataset combination, we perform Bayesian parameter inference using the publicly available Markov Chain Monte Carlo (MCMC) sampler \texttt{emcee} \cite{foremanmackey2013}. The resulting chains are processed with the \texttt{GetDist} package \cite{Lewis2019} to compute the mean values, marginalized posterior distributions, credible intervals, and two-dimensional confidence contours. All free parameters are assigned flat priors over ranges that encompass both the $\Lambda$CDM limit and physically motivated extensions. Since each proposed model introduces exactly two new parameters beyond the base $\Lambda$CDM cosmology—namely $w_2$ for dark matter and one additional parameter ($w_0$, $\alpha$, or $w$) for dark energy—we quantify the improvement in fit relative to $\Lambda$CDM through the difference in minimum chi-squared values, $\Delta\chi^2 = \chi^2_{\Lambda\mathrm{CDM}} - \chi^2_{\rm extended}$. Because the extended models are nested within $\Lambda$CDM, Wilks’ theorem implies that $\Delta\chi^2$ asymptotically follows a $\chi^2$ distribution with $\Delta k = 2$ degrees of freedom. The corresponding $p$-value is $p = 1 - F_{\chi^2_2}(\Delta\chi^2)$, where $F_{\chi^2_2}$ is the cumulative distribution function of the $\chi^2$ distribution with two degrees of freedom, and the equivalent Gaussian significance is obtained by solving $p = \operatorname{erfc}(N_\sigma / \sqrt{2})$, yielding $N_\sigma = \sqrt{2}\,\operatorname{erfc}^{-1}(p)$. To account for model complexity beyond the nested case, we also compute the Akaike Information Criterion, $\mathrm{AIC} = \chi^2_{\min} + 2k$, and the Bayesian Information Criterion, $\mathrm{BIC} = \chi^2_{\min} + k \ln N$, where $k$ is the number of free parameters and $N$ is the number of effective data points in the likelihood. Lower AIC or BIC values indicate preferred models, with differences $\Delta\mathrm{AIC} > 6$ or $\Delta\mathrm{BIC} > 10$ conventionally interpreted as strong evidence against the higher-scoring model. This multi-probe strategy ensures a comprehensive and cross-validated assessment of dynamical dark sector scenarios.

\section{Results and Discussion}\label{sec:results}
    This section presents a comprehensive parameter estimation and model comparison for the three extended dark sector models introduced in Section~\ref{sec:Model} ($w_0 w_2\mathrm{DM}$, $\alpha w_2\mathrm{DM}$, and $ww_2\mathrm{DM}$), using the observational datasets and statistical methodology outlined in Section~\ref{sec:Observational Data}. Table~\ref{all_results} presents the mean and $1\sigma$ credible intervals for the key cosmological parameters in each model. The two-dimensional joint posterior distributions for the five free parameters of each model are shown in Figures~\ref{result_m1_all}, \ref{result_m2_all}, and \ref{result_m3_all} in Appendix~A.

    \begin{table}[h!]
    \begin{center}
    \caption{\label{all_results}The 1$\sigma$ CL fitting results in the $\Lambda$CDM , $w_0 w_2$DM , $\alpha w_2$DM and $w w_2$DM models from the CMB+DESI , CMB+DESI+PP , CMB+DESI+Union3 and CMB+DESI+DESY5 data combinations. Here , $H_0$ is in units of  \unit{km.s^{-1}.Mpc^{-1}}.}
    \resizebox{\linewidth}{!}{ 
    \begin{tabular}{lccccc} 
    \toprule
        \textbf{Model/Dateset} & \textbf{$H_0$} & \textbf{$\Omega_{b}$} & \textbf{$\Omega_{dm}$} & \textbf{$w_2$} & \textbf{$w_0 \ \text{or} \ \alpha \ \text{or} \ w$} \\
    \midrule
        $\Lambda$CDM\\
        CMB+DESI        & $68.67 \pm 0.30$        & $0.0479 \pm 0.0003$          & $0.2505 \pm 0.0035$          & - & - \\
        CMB+DESI+PP     & $68.55 \pm 0.29$        & $0.0480 \pm 0.0003$          & $0.2520 \pm 0.0034$          & - & - \\
        CMB+DESI+Union3 & $68.57 \pm 0.29$        & $0.0480 \pm 0.0003$          & $0.2517 \pm 0.0034$          & - & - \\
        CMB+DESI+DESY5  & $68.44 \pm 0.29$        & $0.0481 \pm 0.0003$          & $0.2532 \pm 0.0034$          & - & - \\
    \midrule        
        $w_0 w_2$DM     &  &  &  &   & $w_0$ \\       
        CMB+DESI        & $68.39^{+0.73}_{-0.74}$ & $0.0479^{+0.0011}_{-0.0010}$ & $0.2525^{+0.0045}_{-0.0046}$ & $-0.0154^{+0.0079}_{-0.0071}$ & $-1.1255^{+0.2790}_{-0.3463}$ \\
        CMB+DESI+PP     & $67.82 \pm 0.62$        & $0.0487^{+0.0009}_{-0.0010}$ & $0.2555 \pm 0.0040$          & $-0.0192^{+0.0080}_{-0.0072}$ & $-1.3536^{+0.2422}_{-0.2627}$ \\
        CMB+DESI+Union3 & $67.84 \pm 0.65$        & $0.0487^{+0.0009}_{-0.0010}$ & $0.2553 \pm 0.0041$          & $-0.0190^{+0.0080}_{-0.0073}$ & $-1.3446^{+0.2459}_{-0.2836}$ \\       
        CMB+DESI+DESY5  & $67.38 \pm 0.59$        & $0.0493^{+0.0008}_{-0.0009}$ & $0.2577 \pm 0.0039$          & $-0.0225^{+0.0082}_{-0.0074}$ & $-1.5117^{+0.2256}_{-0.2286}$ \\        
    \midrule        
        $\alpha w_2$DM &  &  &  &   & $\alpha$ \\        
        CMB+DESI        & $67.51^{+1.19}_{-1.20}$ & $0.0492^{+0.0016}_{-0.0019}$ & $0.2588^{+0.0074}_{-0.0088}$ & $-0.0165 \pm 0.0064$          & $-0.3865^{+0.4322}_{-0.3250}$ \\        
        CMB+DESI+PP     & $67.32^{+0.63}_{-0.62}$ & $0.0494 \pm 0.0009$          & $0.2599^{+0.0048}_{-0.0049}$ & $-0.0170 \pm 0.0058$          & $-0.4304^{+0.1998}_{-0.1985}$ \\        
        CMB+DESI+Union3 & $66.60 \pm 0.83$        & $0.0505 \pm 0.0013$          & $0.2647 \pm 0.0062$          & $-0.0189 \pm 0.0059$          & $-0.6754^{+0.3152}_{-0.2669}$ \\        
        CMB+DESI+DESY5  & $66.58 \pm 0.59$        & $0.0506 \pm 0.0009$          & $0.2647 \pm 0.0047$          & $-0.0191^{+0.0057}_{-0.0058}$ & $-0.6812^{+0.2068}_{-0.2055}$ \\
    \midrule        
        $w w_2$DM       &  &  &  &   & $w$ \\ 
        CMB+DESI        & $67.82 \pm 1.07$        & $0.0487^{+0.0015}_{-0.0016}$ & $0.2566 \pm 0.0071$          & $-0.0158 \pm 0.0067$          & $-0.9628^{+0.0446}_{-0.0448}$ \\        
        CMB+DESI+PP     & $67.36^{+0.62}_{-0.63}$ & $0.0494 \pm 0.0009$          & $0.2595 \pm 0.0047$          & $-0.0174 \pm 0.0060$          & $-0.9435^{+0.0263}_{-0.0262}$ \\        
        CMB+DESI+Union3 & $66.88 \pm 0.78$        & $0.0501 \pm 0.0012$          & $0.2625^{+0.0055}_{-0.0056}$ & $-0.0191^{+0.0062}_{-0.0063}$ & $-0.9236^{+0.0324}_{-0.0327}$ \\        
        CMB+DESI+DESY5  & $66.68 \pm 0.59$        & $0.0504 \pm 0.0009$          & $0.2637^{+0.0045}_{-0.0046}$ & $-0.0199 \pm 0.0060$          & $-0.9152^{+0.0248}_{-0.0247}$ \\
    \bottomrule  
    \end{tabular}
    }
    \end{center}
    \end{table}

\subsection{Evidence for Non-Cold Dark Matter and Robustness Across Datasets}
\begin{figure}[h!]
    \centering
    \begin{minipage}{0.32\linewidth}
		\centerline{\includegraphics[width=\textwidth]{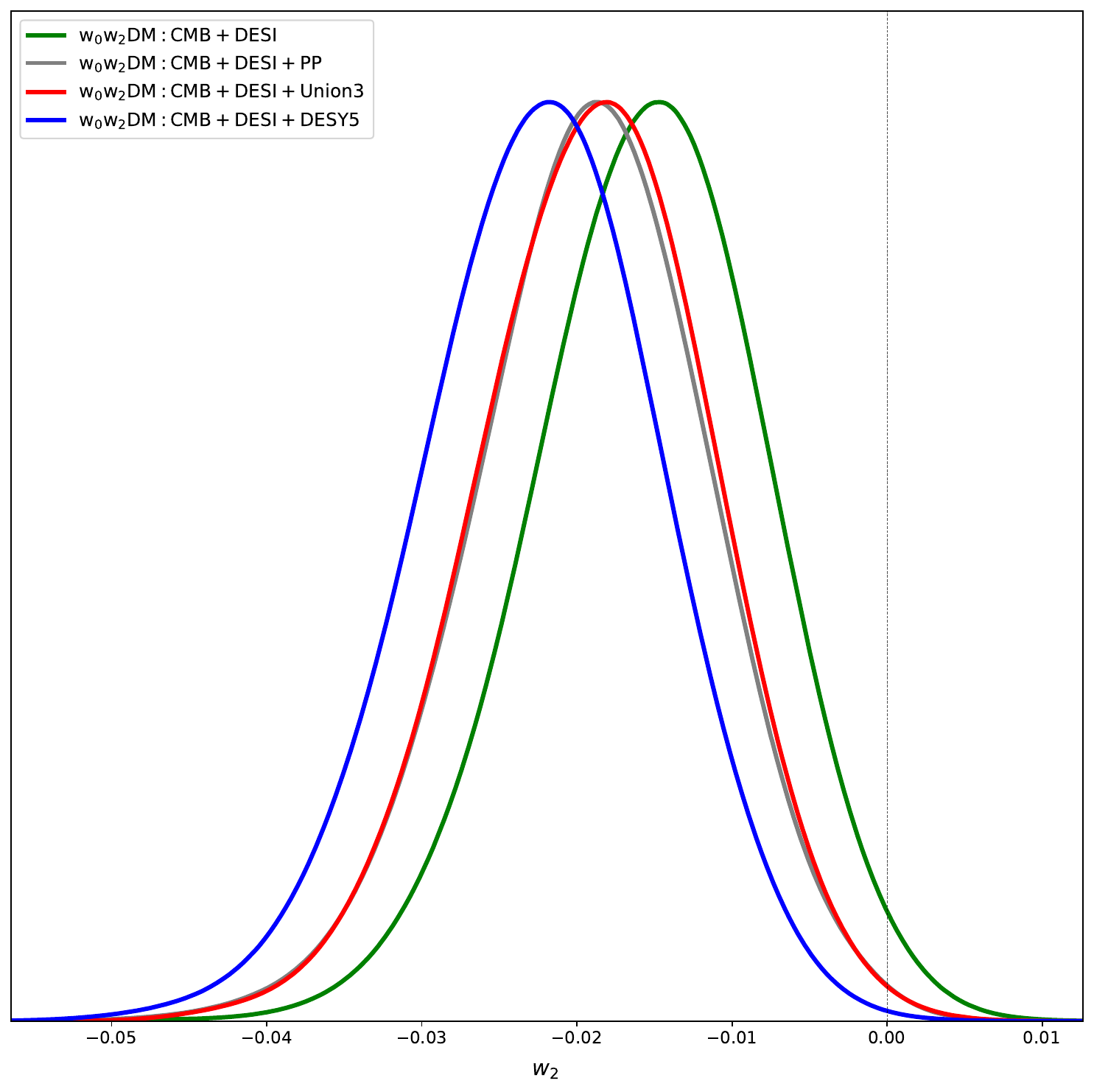}}
	\end{minipage}
	\begin{minipage}{0.32\linewidth}
		\centerline{\includegraphics[width=\textwidth]{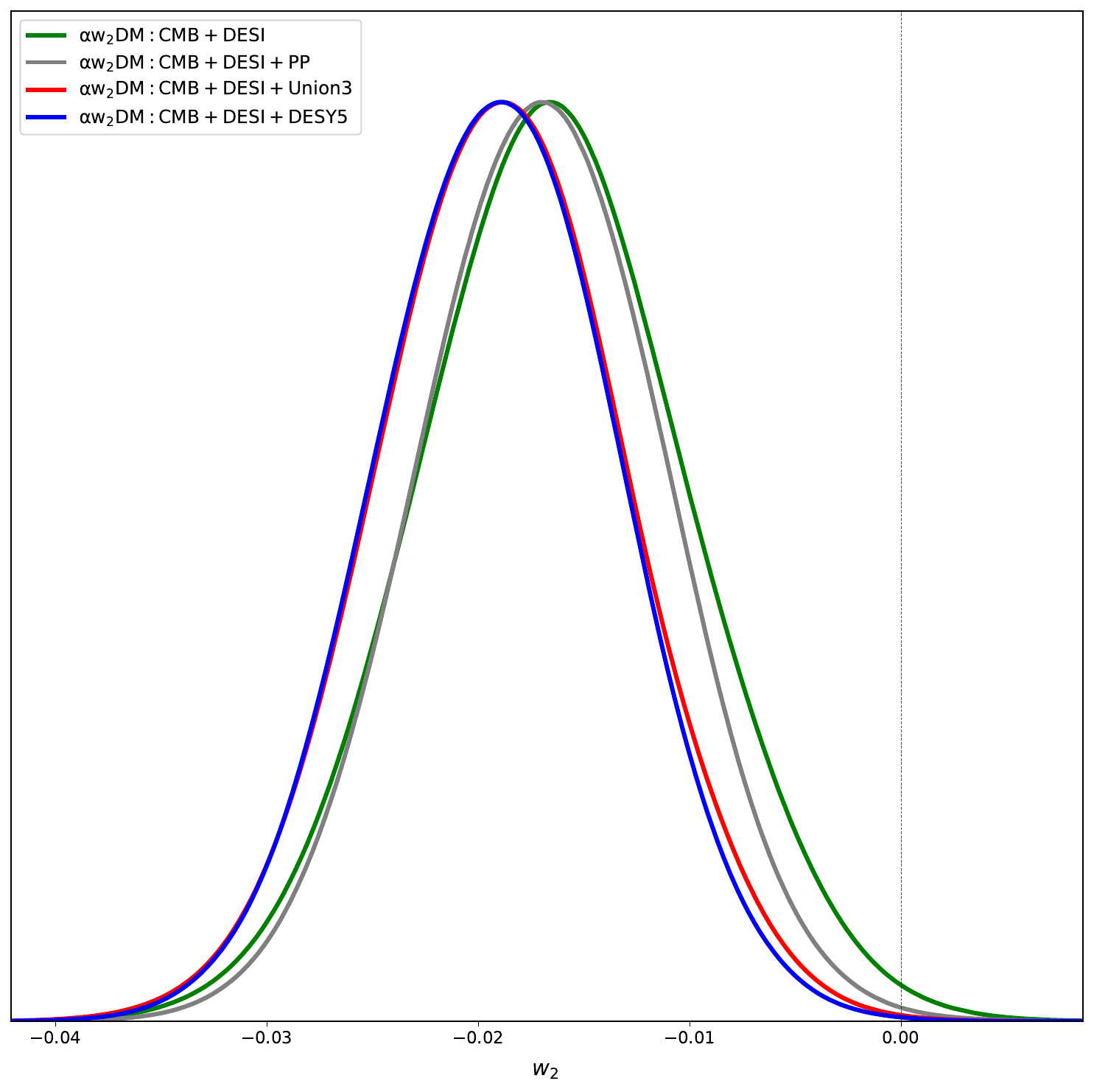}}
	\end{minipage}
	\begin{minipage}{0.32\linewidth}
		\centerline{\includegraphics[width=\textwidth]{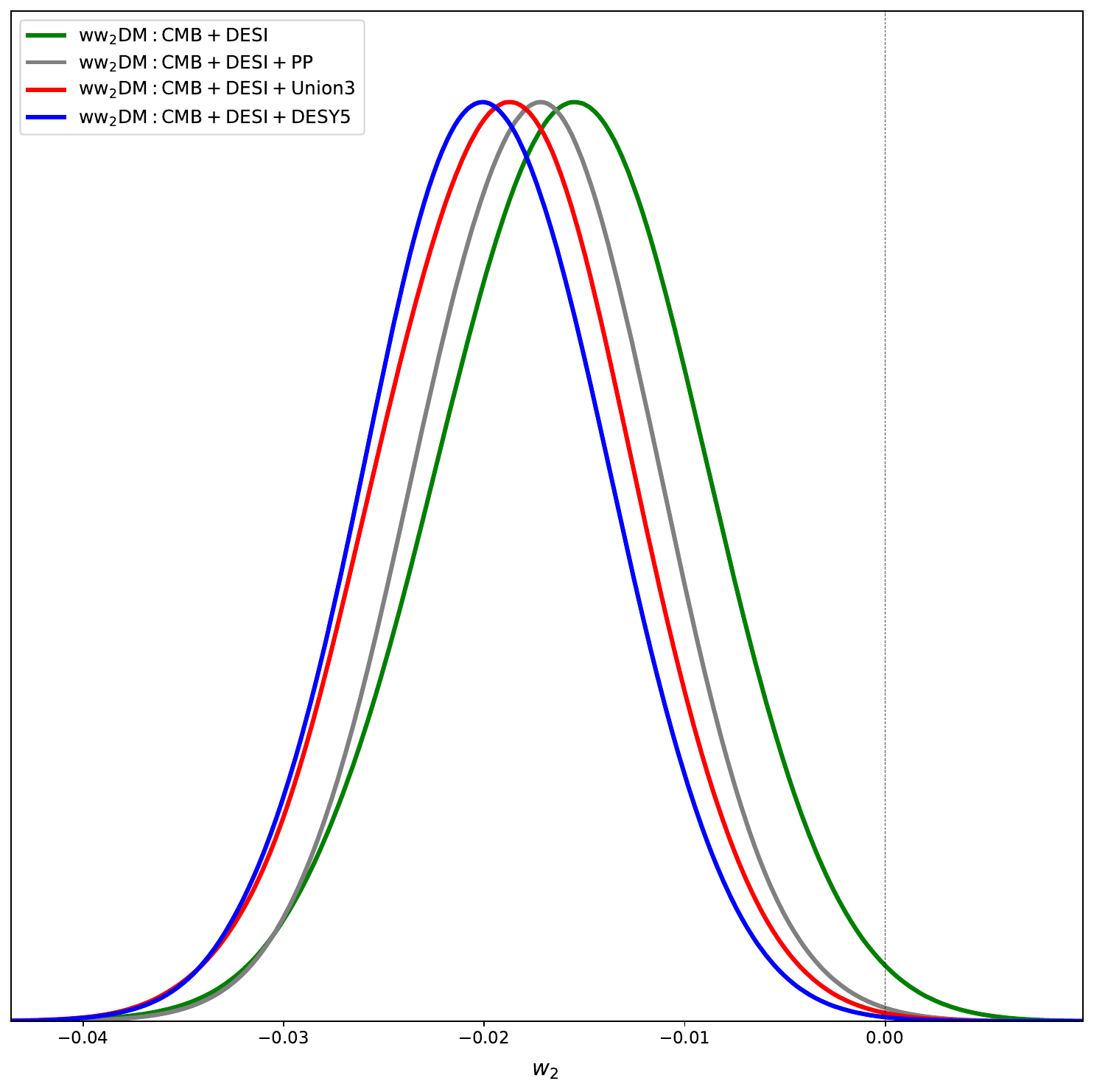}}
    \end{minipage}
	\caption{The one-dimensional marginalized posterior distribution for $w_2$ for $w_0 w_2$DM model(left), $\alpha w_2$DM model(center) and $w w_2$DM model(right), using the data combinations: CMB+DESI, CMB+DESI+PP, CMB+DESI+Union3, and CMB+DESI+DESY5. The vertical dashed lines corresponds to $w_2=0$ .}
	\label{w2_all_models}
    \end{figure}

    As shown in Table~\ref{all_results} and Figure~\ref{w2_all_models}, all three extended dark matter models—$w_0 w_2$DM, $\alpha w_2$DM, and $w w_2$DM—yield consistently negative constraints on the EOS parameter $w_2$ across four independent dataset combinations (CMB+DESI alone and separately combined with the PP, Union3, and DESY5 SN~Ia samples).  
    For the $w_0 w_2$DM model, we find the fitting results of $w_2$ are $-0.0154^{+0.0079}_{-0.0071}$  (CMB+DESI),  $-0.0192^{+0.0080}_{-0.0072}$  (CMB+DESI+PP),  $-0.0190^{+0.0080}_{-0.0073}$  (CMB+DESI+Union3),  and \\$-0.0225^{+0.0082}_{-0.0074}$ (CMB+DESI+DESY5), indicating a preference for a non-zero DM EoS parameter at 1.95$\sigma$, 2.40$\sigma$, 2.38$\sigma$, and 2.74$\sigma$ CL, respectively.
    In the $\alpha w_2$DM model, the constraints tighten to $w_2$ are  $-0.0165 \pm 0.0064$ (CMB+DESI), $-0.0170 \pm 0.0058$ (CMB+DESI+PP), $-0.0189 \pm 0.0059$ (CMB+DESI+Union3), and $-0.0191^{+0.0057}_{-0.0058}$ (CMB+DESI+DESY5), indicating a preference for a non-zero DM EoS parameter at 2.58$\sigma$, 2.93$\sigma$, 3.20$\sigma$, and 3.35$\sigma$ CL, respectively. 
    Similarly, the $w w_2$DM model gives the fitting results of $w_2$ : $-0.0158 \pm 0.0067$ (CMB+DESI), $-0.0174 \pm 0.0060$ (CMB+DESI+PP), $-0.0191^{+0.0062}_{-0.0063}$ (CMB+DESI+Union3), and $-0.0199 \pm 0.0060$ (CMB+DESI+\\DESY5), indicating a preference for a non-zero DM EoS parameter at 2.36$\sigma$, 2.90$\sigma$, 3.08$\sigma$, and 3.32$\sigma$ CL, respectively.
    Critically, almost in every case the inferred $w_2$ is negative and statistically distinct from zero at above 2$\sigma$ confidence. This robust signal present even when using only the purely geometric CMB+DESI combination—strongly indicates that dark matter exhibits a slight negative pressure ($w_{\mathrm{dm}} = w_2 a^2 < 0$), and the significance of this deviation increases when supernova data are included. With the CMB+DESI baseline alone, the improvement over $\Lambda\mathrm{CDM}$ is modest ($\sim 2\sigma$), but it rises to $2.38\sigma$–$3.35\sigma$ once Pantheon+, Union3, or DESY5 are added, so this persistent deviation—robust across parametrizations and data combinations—strongly suggests that the dark matter sector may be more accurately modeled as an effective fluid with non-standard negative pressure, extending beyond the CDM paradigm.

\begin{figure}[h!]
    \centering
    \begin{minipage}{0.32\linewidth}
		\centerline{\includegraphics[width=\textwidth]{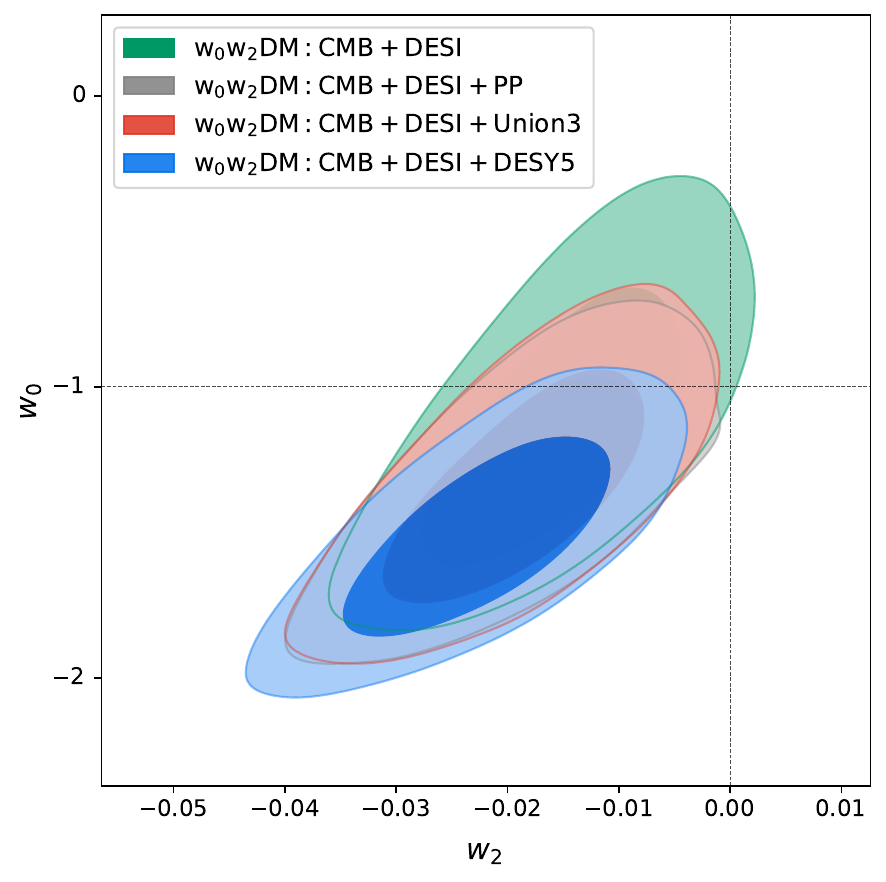}}
	\end{minipage}
	\begin{minipage}{0.32\linewidth}
		\centerline{\includegraphics[width=\textwidth]{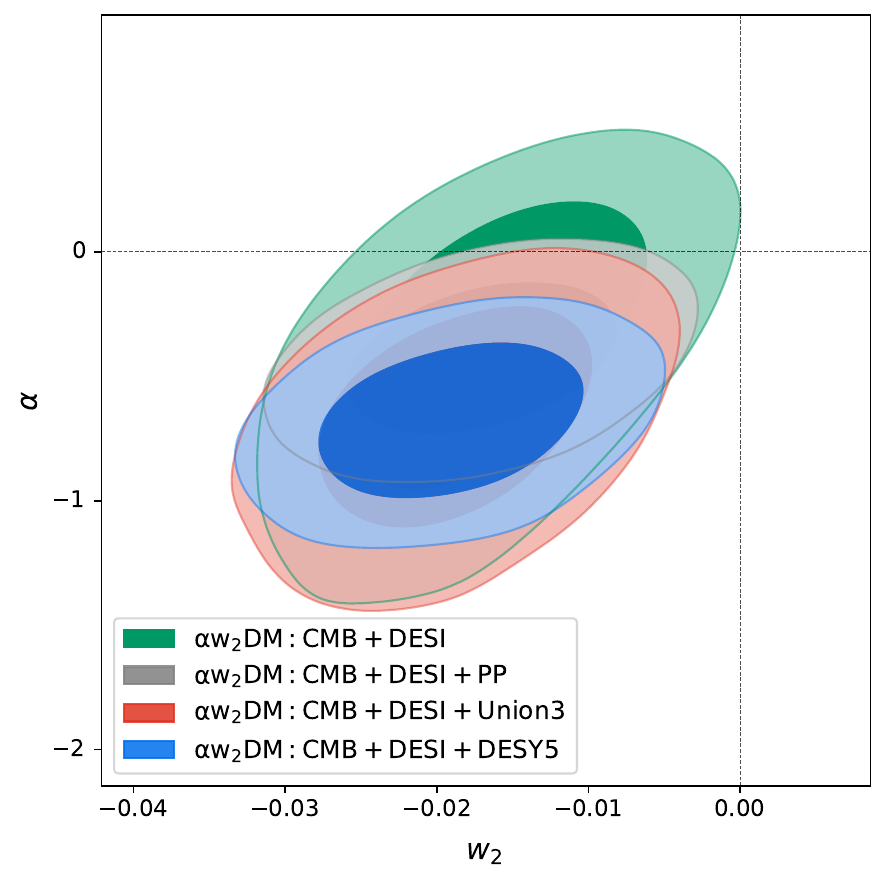}}
	\end{minipage}
	\begin{minipage}{0.32\linewidth}
		\centerline{\includegraphics[width=\textwidth]{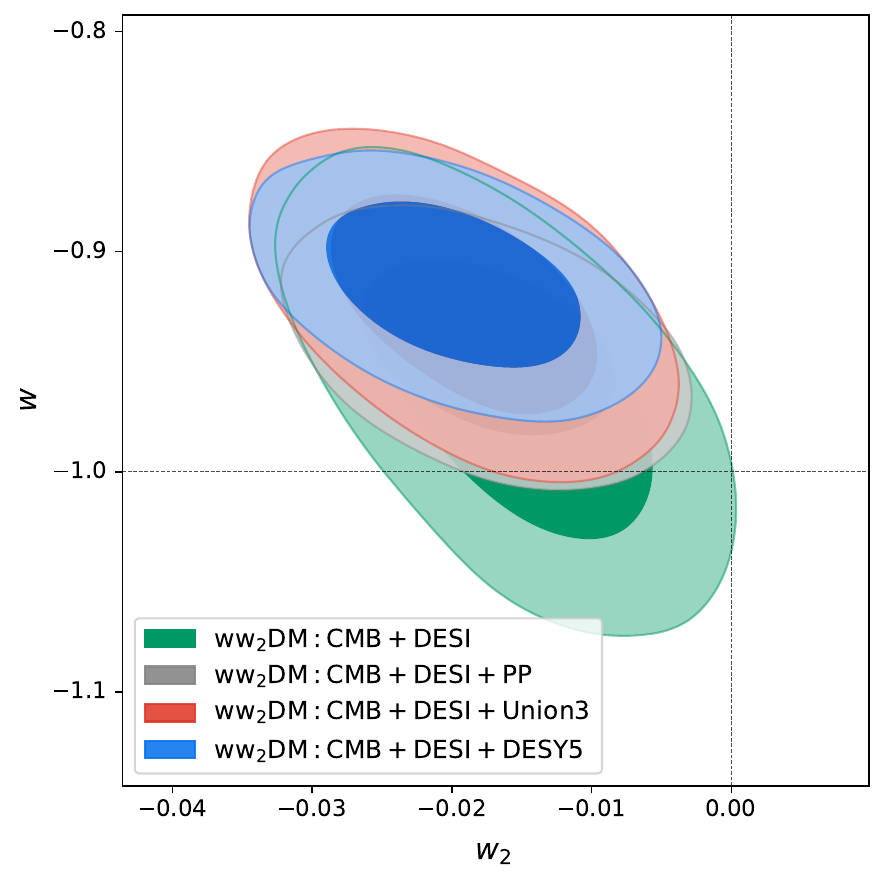}}
    \end{minipage}
	\caption{The two-dimensional joint posterior at 1$\sigma$ and 2$\sigma$ CL of the parameters $w_0$ and $w_2$ (left) in $w w_2$DM model, $\alpha$ and $w_2$(center) in $\alpha w_2$DM model, and $w$ and $w_2$(right) in $w w_2$DM model , using the data combinations: CMB+DESI, CMB+DESI+PP, CMB+DESI+Union3, and CMB+DESI+DESY5. The horizontal dashed line corresponds to $w_0=-1$(left), $\alpha = 0$(center), and $w=-1$(right) , and three vertical dashed lines corresponds to $w_2=0$. }
	\label{2parameters_2d_all_models}
    \end{figure}

    The two-dimensional joint posterior distributions at $1\sigma$ and $2\sigma$ confidence levels for the dark energy and dark matter parameters are shown in Figure~\ref{2parameters_2d_all_models}. In both the $w_0 w_2$DM and $\alpha w_2$DM scenarios (left and center panels), a positive correlation is evident between the dark energy parameter ($w_0$ or $\alpha$) and $w_2$: as dark energy deviates more strongly from the cosmological constant ($w_0 > -1$ or $\alpha > 0$), the preferred value of $w_2$ becomes less negative, and vice versa. In contrast, the $w w_2$DM model (right panel) exhibits a negative correlation between $w$ and $w_2$, bigger value of DE parameter $w$ along with smaller value of DM parameter $w_2$. Despite these differing degeneracy structures, the best-fit values in all three frameworks consistently represent departures from the standard $\Lambda$CDM limit ($w_0 = -1$, $\alpha = 0$, $w = -1$, and $w_2 = 0$). This demonstrates that, although $\Lambda$CDM is formally recoverable as a limiting case, it is disfavored relative to these extended models across multiple independent data combinations. Crucially, all three scenarios robustly yield a negative effective dark matter equation-of-state parameter ($w_2 < 0$), reinforcing the conclusion that the inferred non-cold behavior is not an artifact of a specific dark energy parametrization, but rather a persistent feature of the observational data.

\subsection{Behavior and physical interpretation of dark energy and other cosmological parameters}

\begin{figure}[h!]
    \centering
    \begin{minipage}{0.32\linewidth}
		\centerline{\includegraphics[width=\textwidth]{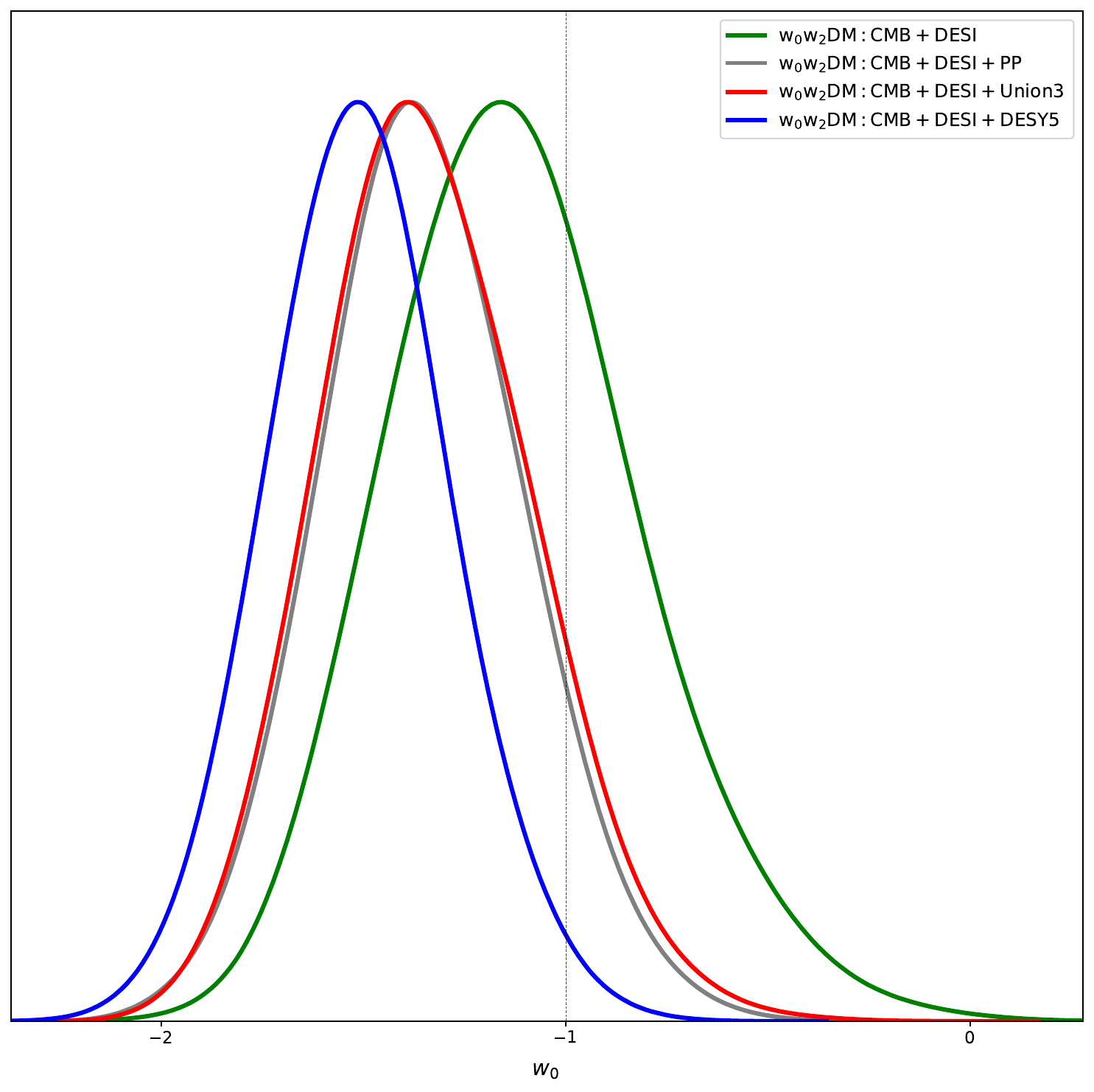}}
	\end{minipage}
	\begin{minipage}{0.32\linewidth}
		\centerline{\includegraphics[width=\textwidth]{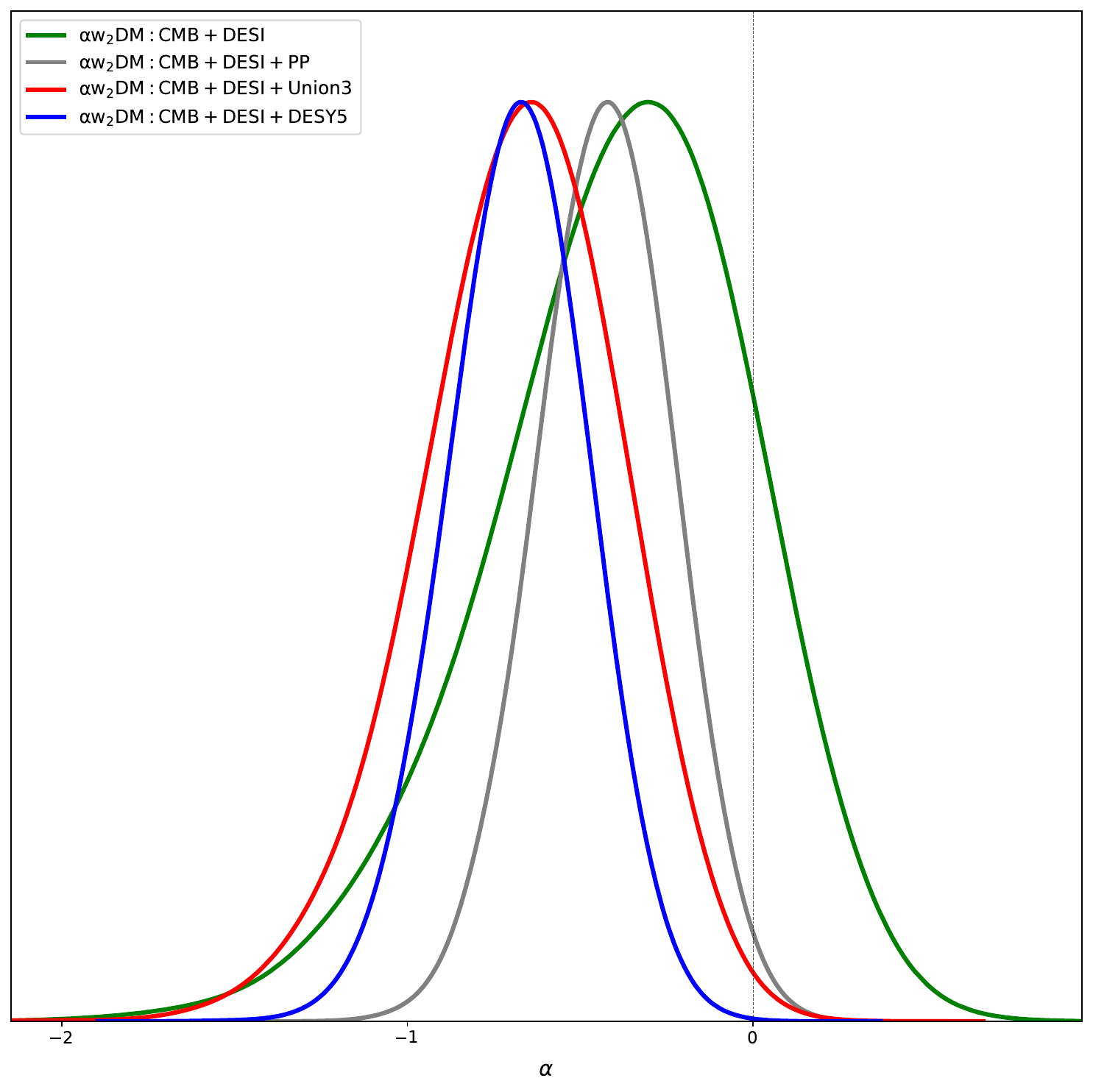}}
	\end{minipage}
	\begin{minipage}{0.32\linewidth}
		\centerline{\includegraphics[width=\textwidth]{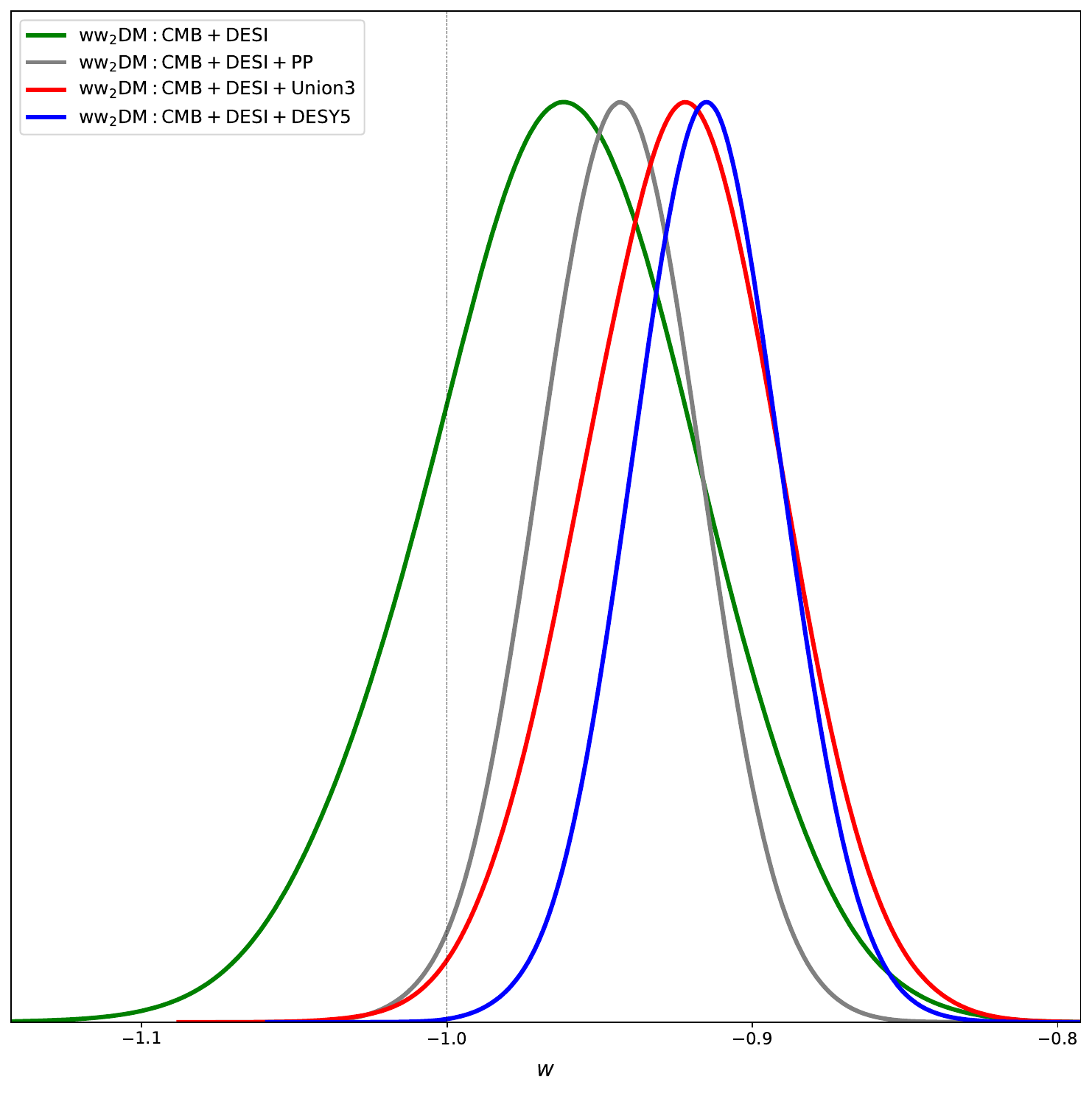}}
    \end{minipage}
	\caption{The one-dimensional marginalized posterior distributions of the parameters $w_0$ (left) in $w w_2$DM model, $\alpha$ (center) in $\alpha w_2$DM model, and $w$ in $w w_2$DM model, using the data combinations: CMB+DESI, CMB+DESI+PP, CMB+DESI+Union3, and CMB+DESI+DESY5. The three vertical dashed lines corresponds to $w_0=-1$(left), $\alpha = 0$(center), and $w=-1$(right).}
	\label{para_de_all_models}
    \end{figure}

The correlations between $w_2$ and the DE parameters ($w_0$, $\alpha$, $w$) shown in Figure~\ref{2parameters_2d_all_models} arise because both dark matter and dark energy contribute to the expansion history. The values of DM and DE parameters deviate from the standard $\Lambda$CDM model (with $w_2=0$ and $w_0=-1$, $\alpha=0$, or $w=-1$)
As shown in Figure~\ref{para_de_all_models}, all three dark energy models exhibit deviations from the cosmological constant $\Lambda$. The thawing scalar field model ($w_0 w_2 \mathrm{DM}$) yields $w_0 < -1$, indicating a possible phantom-like dark energy that may correspond to a scalar field with a specific potential form. In the Modified Emergent Dark Energy model ($\alpha w_2 \mathrm{DM}$), $\alpha < 0$ suggests that dark energy density was suppressed at high redshifts and only emerged recently, differing from the behavior of the original PEDE model\cite{2019A}. The constant-$w$ model ($w w_2 \mathrm{DM}$) gives $w \approx -0.92$ to $-0.96$, representing a mild quintessence-like deviation from $\Lambda$.

These differences between extended models and the $\Lambda$CDM model propagate to key cosmological parameters. We can learn from Table~\ref{all_results} that all extended models yield Hubble constant values between $\sim66.6$ and $68.4$ $\mathrm{km} \ \mathrm{s}^{-1} \ \mathrm{Mpc}^{-1}$, these results are similar with the \textit{Planck} $\Lambda\mathrm{CDM}$ value ($\sim67.4$ \unit{km.s^{-1}.Mpc^{-1}}) \cite{Planck2018-VI} but still below local measurements ($\sim73$ \unit{km.s^{-1}.Mpc^{-1}}) \cite{riess2019large}, so this vividly demonstrates the reality of the Hubble tension. The $\alpha w_2\mathrm{DM}$ model gives the lowest $H_0$ ($\sim66.6$ \unit{km.s^{-1}.Mpc^{-1}}), while $w_0 w_2\mathrm{DM}$ gives the highest. This suggests that the coupling between non-cold dark matter and dynamical dark energy partially absorbs—but does not fully resolve—the $H_0$ tension.

\subsection{Statistical Model Comparison}
    We compare three non-standard dark energy models coupled with non-standard dark matter ($w_0w_2$DM, $\alpha w_2$DM, and $ww_2$DM) against the standard $\Lambda$CDM model. Since all three models reduce to $\Lambda$CDM when their extra parameters are set to specific values ($w_2=0$, and either $w_0=-1$, $\alpha=0$, or $w=-1$), they are nested within $\Lambda$CDM. We therefore use the likelihood ratio test to assess their statistical significance relative to $\Lambda$CDM. Table \ref{chi2_1} presents the minimum $\chi^2$ differences between the $\Lambda$CDM model and the three models considered in this study, defined as $\Delta\chi^2_{\rm min} = \chi^2_{{\rm min},\Lambda{\rm CDM}} - \chi^2_{{\rm min},\rm{extended}}$. It can be seen that, relative to the standard $\Lambda$CDM model, $\Delta\chi^2_{\rm min} > 0$, all three alternative models provide a better fit to the data. Statistical significances of all three extended models ranging from $1.9\sigma$ to $3.7\sigma$. The $\alpha w_2\mathrm{DM}$ model shows the largest improvement—reaching $\sim3.7\sigma$ with CMB+DESI+PP—providing preliminary evidence for physics beyond $\Lambda\mathrm{CDM}$.

    \begin{table}[h!]
    \begin{center}
    \caption{\label{chi2_1}The minimum chi-square differences $\Delta \chi _{min} ^2$ of the $\Lambda$CDM model relative to the $w_0 w_2$DM , $\alpha w_2$CDM, and $w w_2$DM models under CMB+DESI, CMB+DESI+PP, CMB+DESI+Union3, and CMB+DESI+DESY5 data combinations, as well as the significance levels $N_\sigma$ at which these models are preferred over $\Lambda$CDM.}
    \begin{tabular}{lcc} 
    \toprule
        \textbf{Model/Dateset} & \textbf{$\Delta \chi _{min} ^2$} & \textbf{$N_\sigma$} \\
    \midrule       
        $w_0 w_2$DM \\       
        CMB+DESI        &  5.76 & 1.91$\sigma$ \\        
        CMB+DESI+PP     &  9.48 & 2.62$\sigma$ \\        
        CMB+DESI+Union3 &  7.19 & 2.20$\sigma$ \\        
        CMB+DESI+DESY5  &  6.72 & 2.11$\sigma$ \\
    \midrule        
        $\alpha w_2$DM \\        
        CMB+DESI        &  6.22 & 2.01$\sigma$ \\        
        CMB+DESI+PP     & 16.64 & 3.67$\sigma$ \\        
        CMB+DESI+Union3 &  9.77 & 2.67$\sigma$ \\        
        CMB+DESI+DESY5  & 11.00 & 2.87$\sigma$ \\
    \midrule        
        $w w_2$DM \\ 
        CMB+DESI        &  6.18 & 2.00$\sigma$ \\        
        CMB+DESI+PP     & 16.09 & 3.60$\sigma$ \\        
        CMB+DESI+Union3 &  9.65 & 2.65$\sigma$ \\        
        CMB+DESI+DESY5  & 10.58 & 2.80$\sigma$ \\
    \bottomrule
    \end{tabular}
    \end{center}
    \end{table}
    
    Additionally, we compute AIC and BIC values to compare the three new models among themselves, which are not all mutually nested. Since all three extended models ($w_0 w_2$DM, $\alpha w_2$DM, and $w w_2$DM) introduce exactly two additional free parameters beyond the $\Lambda$CDM model, they share the same total number of parameters $k$ and the same effective number of observational date points $N$. Consequently, for a given dataset, the model with the lowest minimum chi-squared ($\chi^2_{\min}$) automatically minimizes both the Akaike Information Criterion (AIC $= \chi^2_{\min} + 2k$) and the Bayesian Information Criterion (BIC $= \chi^2_{\min} + k \ln N$). The differences in information criteria between any two extended models therefore reduce to the difference in their $\chi^2_{\min}$ values:
    \begin{equation}
        \Delta\mathrm{AIC} = \Delta\mathrm{BIC} = \Delta\chi^2_{min}.
    \end{equation}
    Thus, the ranking of the three non-nested models based on AIC or BIC is identical to that based on $\chi^2_{\min}$ alone. Following conventional thresholds~\cite{1995Bayes,2002Model}, a difference $\Delta\mathrm{AIC} = \Delta\mathrm{BIC} > 6$ ($>10$) is interpreted as strong (very strong) evidence against the higher-scoring model.
    
    Table~\ref{chi2_2} shows the pairwise $\Delta\chi^2_{\rm min}$ among these three models themselves. A clear hierarchy emerges when comparing the extended models directly: $\alpha w_2\mathrm{DM}$ consistently outperforms the others, followed by $w w_2\mathrm{DM}$, while $w_0 w_2\mathrm{DM}$ provides the weakest fit. This ranking holds across AIC and BIC, reinforcing its robustness. Crucially, this discrimination is only possible when supernova data are included; with CMB+DESI alone, the models are nearly indistinguishable. The enhanced separation with SN Ia data—particularly with DESY5—highlights how luminosity-distance measurements at redshifts ($0.01 \lesssim z \lesssim 2.3$) complement geometric probes by constraining the late-time expansion history more tightly. Thus, the observed preference for emergent dark energy coupled to non-cold dark matter is not an artifact of a single dataset but a coherent signal strengthened by the synergy of multiple observational pillars.

    \begin{table}[h!]
    \begin{center}
    \caption{\label{chi2_2}The minimum chi-square differences $\Delta \chi _{min} ^2$ of the $w_0 w_2$DM , $\alpha w_2$CDM and $w w_2$DM models under CMB+DESI, CMB+DESI+PP, CMB+DESI+Union3, and CMB+DESI+DESY5 data combinations.}
    \resizebox{\linewidth}{!}{ 
    \begin{tabular}{lcccc} 
    \toprule
        \textbf{Model} & \textbf{CMB+DESI} & \textbf{CMB+DESI+DESY5} & \textbf{CMB+DESI+PP} & \textbf{CMB+DESI+Union3} \\
    \midrule       
    $\chi^2_{min}(w_0 w_2 DM)-\chi^2_{min}(w w_2 DM)$ & 0.42 & 6.61 & 2.46 & 3.85 \\
    \midrule        
    $\chi^2_{min}(w_0 w_2 DM)-\chi^2_{min}(\alpha w_2 DM)$ & 0.45 & 7.17 & 2.58 & 4.28 \\
    \midrule        
    $\chi^2_{min}(w w_2DM)-\chi^2_{min}(\alpha w_2 DM)$ & 0.03 & 0.55 & 0.12 & 0.42 \\
    \bottomrule
    \end{tabular}
    }
    \end{center}
    \end{table}

\section{Conclusion}\label{sec:conclusion}

    In this study, we systematically investigated a non-standard dark matter (NSDM) model parameterized by $w_{\mathrm{dm}} = w_2 a^2$ and coupled it with three different single-parameter dark energy (DE) models: the thawing scalar field model ($w_0 w_2\mathrm{DM}$), the Modified Emergent Dark Energy (MEDE) model ($\alpha w_2\mathrm{DM}$), and the constant $w$ model ($ww_2\mathrm{DM}$). The analyses employ the CMB and DESI BAO datasets, combined with three different SN~Ia datasets: PP, Union3, and DES-Y5. The proposed NSDM model, combined with the three aforementioned DE scenarios, leads to three specific models referred to as $w_0 w_2$DM, $\alpha w_2$DM, and $ww_2$DM, respectively. The results show that for all three models---$w_0 w_2$DM, $\alpha w_2$DM, and $ww_2$DM---and across all dataset combinations (CMB+DESI, CMB+DESI+PP, CMB+DESI+Union3, and CMB+DESI+DESY5), the parameter $w_2$ exhibits a preference for negative mean values. This strongly indicates that the effective dark matter exhibits a "non-cold" property independent of the specific form of the coupled dark energy model, demonstrating considerable robustness. This robust preference indicates a effective dark matter fluid exhibiting a mild negative pressure. Crucially, this conclusion holds irrespective of the assumed dark energy parametrization, underscoring the model-independent nature of the inferred effective dark matter property.

    Observational data tend to favor dark energy being dynamical rather than a strict cosmological constant. Among the three models, the coupled model of Modified Emergent Dark Energy (MEDE) and NSDM ---$\alpha w_2\mathrm{DM}$--- receives the strongest data support, with its goodness-of-fit improvement relative to $\Lambda\mathrm{CDM}$ reaching up to approximately $3.7\sigma$. This suggests that a dark energy component that dynamically emerges in the late universe, combined with a slightly non-cold dark matter, can more harmoniously describe the current cosmological observations. The introduction of NSDM and dynamical DE alters the expansion history of the universe. The estimated Hubble constant $H_0$ from these models lies between the early universe inferences and some local direct measurements. Although this does not completely resolve the Hubble tension, it demonstrates the potential to reconcile observational discrepancies through the synergistic evolution of dark matter and dark energy properties.

    In summary, this study finds that models allowing for dynamical dark energy are statistically preferred, while novelly revealing the non-cold nature of dark matter. The consistent preference for $w_2 < 0$ alongside deviations from a cosmological constant highlights the flexibility of extended dark sector models in describing current observations. These findings also hint at possible deviations in other cosmological parameters beyond $\Lambda$CDM, warranting further investigation with upcoming observational probes (such as LSST, Euclid and upcoming CSST etc.) to test such extensions and uncover their possible underlying physical origins. We
will take efforts to continue these investigating lines.

\section*{Acknowledgement}
The present work is partly supported by NSFC and we appreciate Dr. Y-H Yao for lots of fruitful discussions and long time communications.

\begin{appendices}\label{appendix}
    \section{Cosmological constraints on all five parameters}
    \begin{figure}[h!]
    \centering
		\includegraphics[width=\textwidth]{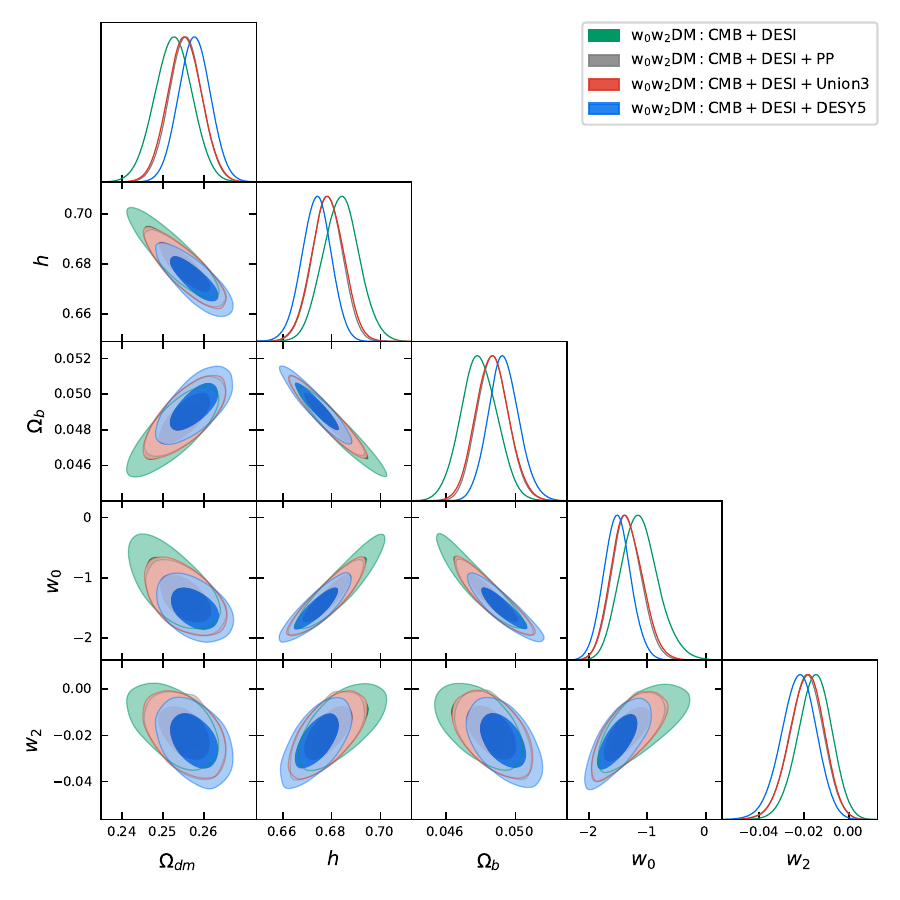}
	\caption{The triangular plot of the fitting results for the $w_0 w_2$DM model, using the data combinations: CMB+DESI, CMB+DESI+PP, CMB+DESI+Union3, and CMB+DESI+DESY5.}
	\label{result_m1_all}
    \end{figure}

    \begin{figure}[h!]
    \centering
		\includegraphics[width=\textwidth]{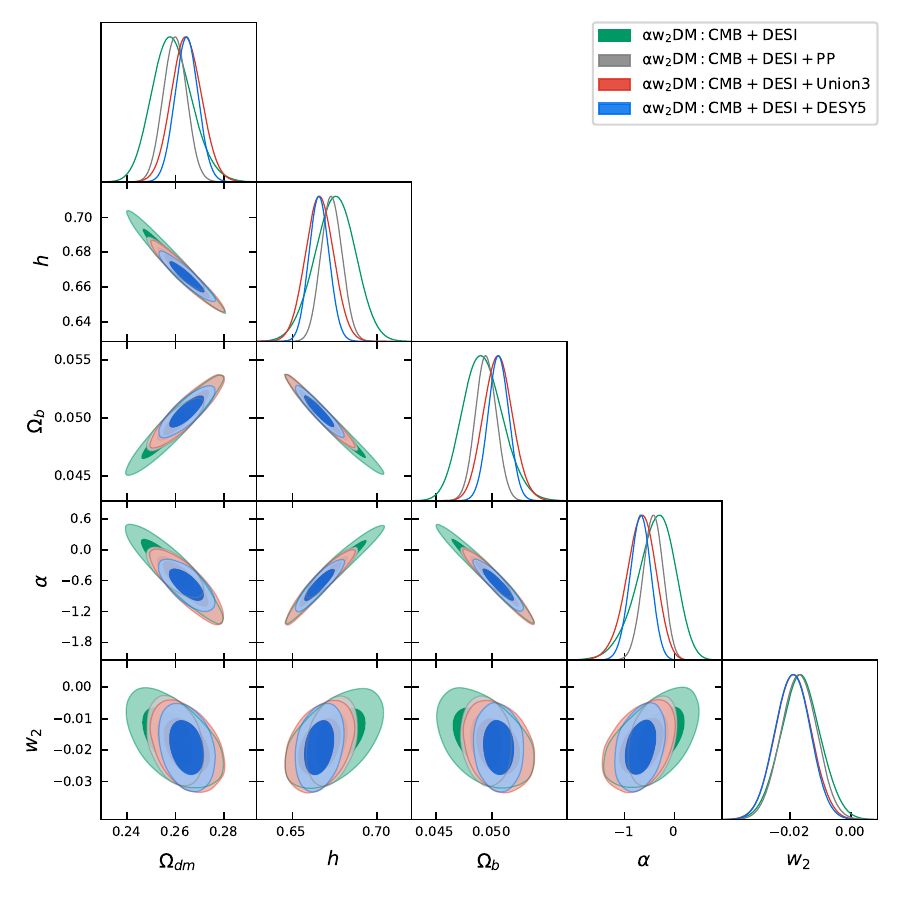}
	\caption{The triangular plot of the fitting results for the $\alpha w_2$DM model, using the data combinations: CMB+DESI, CMB+DESI+PP, CMB+DESI+Union3, and CMB+DESI+DESY5.}
	\label{result_m2_all}
    \end{figure}

    \begin{figure}[h!]
    \centering
		\includegraphics[width=\textwidth]{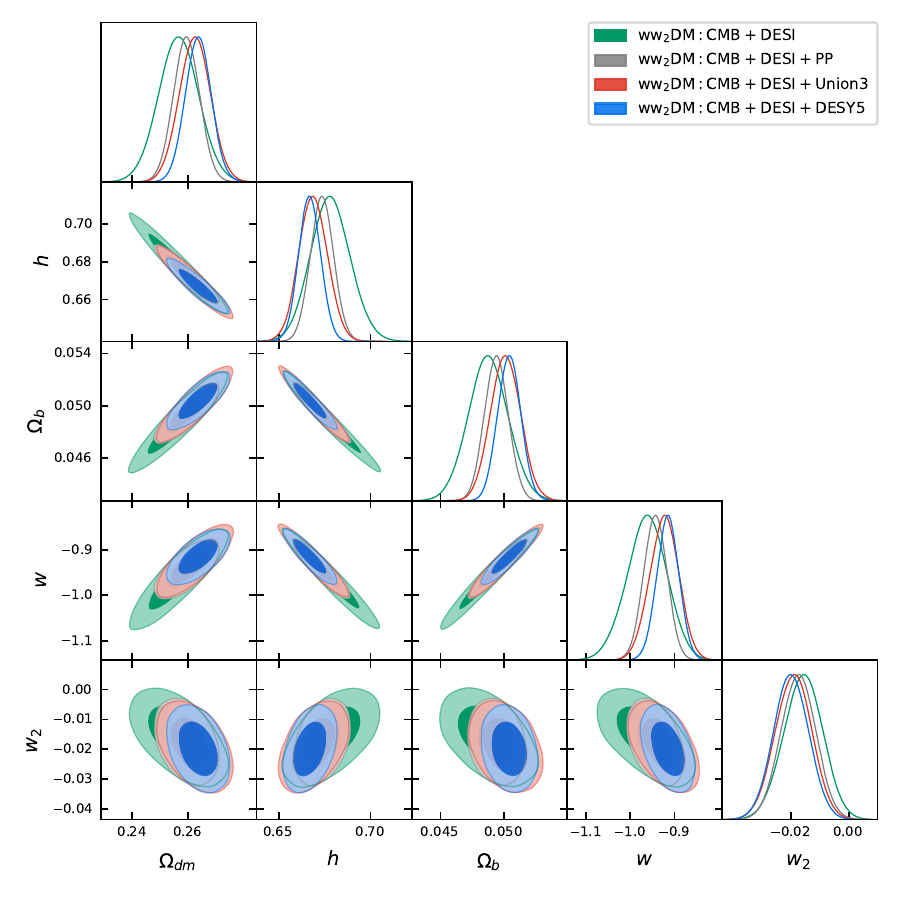}
	\caption{The triangular plot of the fitting results for the $w w_2$DM model, using the data combinations: CMB+DESI, CMB+DESI+PP, CMB+DESI+Union3, and CMB+DESI+DESY5.}
	\label{result_m3_all}
    \end{figure}

\end{appendices}

\FloatBarrier 

\bibliographystyle{unsrt}
\bibliography{test_ref}
    
\end{document}